\documentclass[12pt, twoside, epsf]{article}

\usepackage{latexsym}
\usepackage{amsmath}
\usepackage{amssymb}
\usepackage{amsfonts}
\usepackage{mathrsfs}
\usepackage{ifthen}
\usepackage{times}
\usepackage{graphicx}
\usepackage{array}
\usepackage{flafter}
\usepackage{fullpage}

\newcommand{\ex}[1]{example~\ref{ex#1}}

\newtheorem{ExampleDef}{Example}[section]

\newcommand{\Example}[3]{
  \begin{list}{}{
      \setlength{\leftmargin}{1em}}     
    \item                               
    \small                              
    \begin{ExampleDef} \rm              
      {\bf \hspace{-1ex}: #1}           
      #2                                
      \hfill {\large \boldmath $\Box$}  
      \label{ex#3}                      
    \end{ExampleDef}
  \end{list}}

\begin{document}

\begin{center}
{\Large \bf Contraction Analysis of Nonlinear Distributed Systems \par}
\vspace{1.5em}
{\large Winfried Lohmiller and Jean-Jacques E. Slotine \par}
{Nonlinear Systems Laboratory \\
Massachusetts Institute of Technology \\
Cambridge, Massachusetts, 02139, USA\\
{\sl wslohmil@mit.edu, jjs@mit.edu} \par}
\vspace{3em}
\end{center}

\begin{abstract}
Contraction theory is a recently developed dynamic analysis and
nonlinear control system design tool based on an exact differential
analysis of convergence. This paper extends contraction theory to
local and global stability analysis of important classes of nonlinear
distributed dynamics, such as convection-diffusion-reaction processes,
Lagrangian and Hamilton-Jacobi dynamics, and optimal controllers and
observers.

The Hamilton-Jacobi-Bellman controller and a similar optimal nonlinear
observer design are studied. Explicit stability conditions are given,
which extend the well-known conditions on controllability and
observability Grammians for linear time-varying systems.

Stability of the Hamilton-Jacobi dynamics is assessed by evaluating
the Hessian of the system state along system trajectories. In contrast
to stability proofs based on energy dissipation, this principle allows
to conclude on stability of energy-based systems that are excited by
time-varying inputs. In this context, contraction can be regarded as
describing new variational conservation laws and the stability of
entropy producing processes.

\end{abstract}

\newpage

\section{Introduction}

This paper shows how the tools of contraction theory can contribute to
the stability and convergence analysis of nonlinear partial
differential processes \cite{Clay,Evan,PLL,Kristic}.  Contraction
theory is a recently developed analysis and control system design tool
based on an exact differential analysis of convergence. While
differential approximation is the basis of all linearized stability
analysis, what is new in contraction theory is that differential
stability analysis can be made {\it exact}, and in turn yield global
results on the nonlinear system \cite{Lohm0,Lohm1}.

Specifically, the paper considers stability and convergence analysis
of vector nonlinear diffusion processes and Hamiltonian dynamics, and
more broadly of nonlinear partial differential equations in the
general form
\begin{equation}
\frac{\partial \phi_i}{\partial t} + h_i (\Phi, \nabla \phi_i,
\vec{x}, t) = \nabla \cdot {\bf G}_i ( \nabla \Phi, \vec{x}, t) \label{eq:pde}
\end{equation}
where $\Phi$ is the $n$-dimensional continuous state vector (whose
components $\phi_i $ consist, for instance, of chemical concentrations
or mechanical velocities), and $\vec{x}$ is the $m$-dimensional vector
of underlying coordinates. Specific application examples are then
discussed, including distributed chemical processes, classical fluid
systems such as Bernoulli and Navier-Stokes dynamics, Lagrangian
systems, Hamilton-Jacobi dynamics, and Hamilton-Jacobi-Bellman designs
of optimal controllers or observers \cite{Aris, Bryson, Evan}.  We
also show that approximations of such processes using basis functions
are themselves contracting, regardless of the coarseness of the
approximation.

Basically, a nonlinear dynamic system is called contracting if initial
conditions or temporary disturbances are forgotten exponentially fast,
i.e., if trajectories of the perturbed system return to their nominal
behavior with an exponential convergence rate. In many physical
systems this behavior can be interpreted as entropy producing $-$ when
the state corresponds to a physical quantity as velocity, temperature
or a chemical concentration, the variation between actual and nominal
behavior uses up available energy. As shown in \cite{Lohm1} (to which
the reader is referred for more details), it turns out that, for
ordinary differential equations, relatively simple conditions can be
given for this stability-like property to be verified, and furthermore
that this property is preserved through basic system combinations,
such as parallel combinations, series or hierarchies, and specific
feedback combinations.

In order to extend these results to the distributed case
(\ref{eq:pde}), we will regard any distributed quantity in the
following as a limit of a spatial discretization. Accordingly the
$\nabla$-operator will be regarded as a limit of an explicit
differentation and the $\int$-operator as a limit of a sum. Particular
instances of these extensions were obtained in \cite{Lohm2,Lohm3}, and
in \cite{Kris, Egel} using operator notation, which we adopt here. In
the following $\ \nabla \cdot\ $ corresponds to a inner product with
the $\nabla$ operator, while $\nabla$ represents an outer product with
the $\nabla$ operator.

Section \ref{firstorderpde} analyzes the contraction behavior of
nonlinear first-order partial differential equations and their
numerical solution, and discusses specific applications.  Section
\ref{higherorderpde} studies similarly second-order partial
differential equations. Brief concluding remarks are offered in
section~\ref{concluding}.

\section{First-order partial differential equations} \label{firstorderpde}
This section concentrates on first-order nonlinear partial
differential equations on a continuum $V$, such as
e.g. Hamilton-Jacobi equations, in the general form
\begin{equation}
\frac{\partial{\phi_i}}{\partial t} + h_i (\Phi, \nabla \phi_i, \vec{x},
t) = 0 \label{eq:firstorderdynamics}
\end{equation}

Motivated by contraction analysis of ordinary systems in \cite{Lohm1},
consider two neighboring solutions of (\ref{eq:firstorderdynamics})
{\it at fixed time $t$}, and the virtual displacement $\delta
\Phi$. This leads to the virtual dynamics
\begin{equation}
\frac{\partial}{\partial t}\ \delta \phi_i + \frac{\partial h_i}{\partial
\Phi}\ \delta \Phi + \frac{\partial h_i}{\partial \nabla \phi_i}\ \delta
\nabla \phi_i = 0 \label{eq:deltaphidot}
\end{equation}
In most physical and technical applications, such as Bernoulli,
Navier-Stokes and Hamilton-Jacobi systems, the term $\frac{\partial
h_i}{\partial \nabla \phi_i}$ corresponds to an underlying velocity
field of the quantity $\phi_i$ in (\ref{eq:firstorderdynamics}), and
so we will refer to this term as the flow ``velocity''.

Existence and uniqueness questions in nonlinear partial differential
equations are notoriously difficult \cite{Evan,Clay,PLL}.  Because the
solution of (\ref{eq:pde}) or (\ref{eq:firstorderdynamics}) needs not
be differentiable with respect to $\vec{x}$, $\nabla \phi\ $ has to be
carefully defined.  We choose to define $\nabla \phi_i\ $ causally
along flow lines, i.e., so that it corresponds, in the sense of
distributions, to the limit of an explicit backward differentiation
against the corresponding velocity component $\frac{\partial
h_i}{\partial \nabla \phi_i}$ (in the case that a component is zero,
we can choose the direction of the differentiation arbitrarily without
affecting (\ref{eq:deltaphidot})). Accordingly we assume $\phi_i$ to
be given at all {\it inflowing boundaries}, i.e., parts of the
continuum boundary $\partial V$ where $\frac{\partial h_i}{\partial
\nabla \phi_i}\ {\bf n} < 0$, with ${\bf n}$ the unit vector along the
outward normal. Interpreting (\ref{eq:firstorderdynamics}) as an
ordinary infinitesimal discretization in space can then lead to a natural
existence condition on the solution of (\ref{eq:pde}).

In this section, we derive two main results, one for a fixed
continuum, and another for a continuum moving and deforming according
to a predefined velocity field in $\vec{x}$ space. Applications of the
first result includes transport laws and Bernoulli dynamics, while
applications of the second includes Hamilton-Jacobi dynamics and
optimal controller/observer design.

\subsection{First-order p.d.e.'s on a fixed continuum}

Let us analyze the contraction behavior of (\ref{eq:deltaphidot}) on a
fixed region $V$. We first derive the main result using partial
integration, under the assumption that $\delta \phi_i$ and
$\frac{\partial h_i}{\partial \nabla \phi_i}$ are continuously
differentiable, before extending it to possibly discontinuous $\delta
\phi_i$ and $\frac{\partial h_i}{\partial \nabla \phi_i}$ using a more
technical proof.

We start with the flux equation, for a continuum $V$
of boundary $\partial V$,
$$ 
\int_V \nabla \cdot \left( \delta \phi_i \frac{\partial h_i}{\partial
\nabla \phi_i} \delta \phi_i \right) dV = \int_{\partial V} \delta \phi_i
\frac{\partial h_i}{\partial \nabla \phi_i} \ {\bf n}\ \delta \phi_i \ d
\partial V
$$ 
where ${\bf n}$ is the unit vector along the outward normal of
$\partial V$, and $\frac{\partial h_i}{\partial \nabla \phi_i} \ {\bf
n}$ the corresponding outward flux. Assuming that $\phi_i$ is given at
all inflowing boundaries, and using
$$
\nabla \cdot \left( \delta \phi_i \frac{\partial h_i}{\partial \nabla \phi_i}
\delta \phi_i \right) = 2 \delta \phi_i \frac{\partial h_i}{\partial \nabla
\phi_i} \delta \nabla \phi_i + \delta \phi_i \nabla \cdot \frac{\partial
h_i}{\partial \nabla \phi_i} \delta \phi_i
$$
the time derivative of $\int_V \delta \phi_i \delta \phi_i \ dV$ can be
computed with partial integration as
\begin{eqnarray*}
\frac{1}{2}\frac{d}{dt} \int_V \delta \phi_i \delta \phi_i dV & = & \int_V
\delta \phi_i \ \frac{\partial}{\partial t}\ \delta \phi_i \ dV \ = \int_V
\delta \phi_i \ \left( - \frac{\partial h_i}{\partial \Phi}\ \delta \Phi -
\frac{\partial h_i}{\partial \nabla \phi_i} \delta \nabla \phi_i \right) \ dV
\\ & = & \int_V - \delta \phi_i \ \frac{\partial h_i}{\partial \Phi} \delta
\Phi + \delta \phi_i \frac{\nabla}{2} \cdot \frac{\partial h_i}{\partial
\nabla \phi_i} \delta \phi_i \ dV - \ \frac{1}{2} \
\int_{\partial V} \delta \phi_i \frac{\partial h_i}{\partial \nabla
\phi_i} \ {\bf n}\ \delta \phi_i \ d \partial V
\end{eqnarray*}
This yields the bound and main result
\begin{equation}
\frac{1}{2} \frac{d}{dt} \int_V \delta \Phi^T \delta \Phi \
dV \ \le \ \int_V \delta \Phi^T \left(-
\frac{\partial {\bf h}}{\partial \Phi}\ + \frac{\nabla}{2} \cdot
\frac{\partial {\bf h}}{\partial \nabla \Phi} \right) \delta \Phi \
dV \ \le \ \lambda_{V}\ \int_V \delta \Phi^T \delta
\Phi \ dV \label{eq:firstordershrink}
\end{equation}
where $\lambda_{V}$ denotes the largest eigenvalue of the symmetric part of $-
\frac{\partial {\bf h}}{\partial \Phi}\ + \frac{\nabla}{2} \cdot
\frac{\partial {\bf h}}{\partial \nabla \Phi}\ $ .

Assume now that $\lambda_{V}$ is uniformly strictly negative (i.e., $\
\exists \ \beta > 0,\ \forall {\Phi},\ \forall t \geq 0, \ \lambda_{V}
\leq - \beta < 0$) then from (\ref{eq:firstordershrink}) any $\int_V
\delta \Phi^T \delta \Phi \ dV$ converges exponentially to zero. As in
contraction analysis for ordinary differential equations, this
immediately implies, by path integration, that any finite difference
between two arbitrary solutions converges exponentially to zero. Thus,
as in stable linear time-invariant systems, the initial conditions are
exponentially ``forgotten.''

In the case that $\delta \phi_i$ or $\frac{\partial h_i}{\partial
\nabla_j \phi_i}$ is not continuously differentiable, inequality
(\ref{eq:firstordershrink}) can still be obtained using a more
technical proof, as discussed in the Appendix.

The argumentation can be extended to the following cases:
\begin{itemize}
\item Consider again the dynamics (\ref{eq:firstorderdynamics}) with
an upper bounded $-\frac{\partial {\bf h}}{\partial \Phi}\ $ and a
given initial condition ${\Phi}_o$. This implies that $\delta \Phi^T
\delta \Phi$ remains equal to zero $\forall t \ge 0$, which implies
uniqueness of the solution $\forall t \ge 0$.

\item Apply now to system (\ref{eq:firstorderdynamics}) a set of {\it
linear} constraints in $\Phi$ or $\nabla \Phi$.  Such a case may
describe, for instance, mechanical systems with kinematic constraints,
incompressible fluid flows, and chemical systems in partial
equilibrium. The constrained dynamic equations take the form
$$
  \frac{\partial \phi_i}{\partial t} + h_i(\Phi, \nabla \phi_i,
  \vec{x}, t) + p_i = 0
$$ 
where the constraint terms ${\bf p}(\Phi, \nabla \Phi, \vec{x}, t)$
are orthogonal to the constraint plane. Interpreting the above as
an ordinary limit of a spatial discretization, we can conclude from
\cite{Lohm1} that contraction is preserved.
\end{itemize}

In summary,

\newtheorem{theorem}{Theorem}
\begin{theorem}
Consider the dynamics 
$$
\frac{\partial{\phi_i}}{\partial t} + h_i (\Phi,
\nabla \phi_i, \vec{x}, t) + p_i = 0
$$
with time $t$, $m$-dimensional coordinates $\vec{x}$, $n$-dimensional state
vector $\Phi$, $p$ constraint terms ${\bf p}$ orthogonal to $p$ linear
algebraic constraints in $\Phi$ and $\nabla \Phi$, and $\nabla$ in line $i$
defined as the limit of a backward differentation against the velocity
$\frac{\partial h_i}{\partial \nabla \phi_i}$.

For given $\phi_i$ over all inflowing boundaries $\partial V$
(i.e. $\frac{\partial h_i}{\partial \nabla \phi_i} \ {\bf n} < 0$) and
uniformly negative definite $-\frac{\partial {\bf h}}{\partial \Phi}\
+\ \frac{1}{2}\ \nabla \cdot \frac{\partial {\bf h}}{\partial \nabla
\Phi}\ $ the system converges exponentially to a single solution,
independent of initial conditions. The convergence rate is $|
\lambda_{V} |$, where $\lambda_{V}$ is the largest eigenvalue of the
symmetric part of $-\frac{\partial {\bf h}}{\partial \Phi}\ +\
\frac{1}{2}\ \nabla \cdot \frac{\partial {\bf h}}{\partial \nabla
\Phi}$.

In the autonomous case ( $h_i = h_i(\Phi, \nabla \phi_i,
\vec{x})$, and constant boundary conditions and algebraic constraints)
and under the same conditions, the system converges exponentially to
a unique steady-state $\frac{\partial \Phi}{\partial t} = {\bf 0}$.
\label{th:firstorderpde}
\end{theorem}

We will call a system contracting if Theorem \ref{th:firstorderpde}
applies.  Similarly, we call the system semi-contracting for
$-\frac{\partial {\bf h}}{\partial \Phi}\ +\ \frac{1}{2}\ \nabla \cdot
\frac{\partial {\bf h}}{\partial \nabla \Phi} \le 0$ and indifferent
for $-\frac{\partial {\bf h}}{\partial \Phi}\ +\ \frac{1}{2}\ \nabla
\cdot \frac{\partial {\bf h}}{\partial \nabla \Phi} = 0$.




More general local versions of the above theorem can also be derived as in
\cite{Lohm1}. Note that as the examples illustrate, contraction behavior can
be interpreted physically as entropy producing, since $\delta \phi$ could have
been used to spend available energy instead, if the state corresponds to a
physical quantity such as velocity, temperature, or chemical concentration.

%

\Example{}{Consider the conservation of a quantity $\phi$ in an
$m$-dimensional continuum $V$
$$ 
\frac{\partial \phi}{\partial t} + \vec{v}(\vec{x}, t) \ \nabla \phi = 0
$$ 
with $\phi$ given at all inflowing boundaries. Computing 
$$
- \frac{\partial h}{\partial \phi}\ +\ \frac{1}{2}\ \nabla \cdot \frac{\partial
h}{\partial \nabla \phi}\ = \ \frac{1}{2}\nabla \cdot \vec{v}
$$ 
we conclude with Theorem \ref{th:firstorderpde} on contraction behavior for a
compressing flow field with $\nabla \cdot \ \vec{v} \ \le \ 0$, with minimal
convergence rate $|\frac{\nabla \cdot \vec{v}}{2}|$. 

Consider another typical conservation of $\phi$ in an $m$-dimensional
continuum $V$
$$ 
\frac{\partial \phi}{\partial t} \ + \ \nabla \cdot (\phi
\vec{v}(\vec{x}, t)) = 0
$$ with $\phi$ given at all inflowing boundaries and equilibrium
points. Similarly to the above we can conclude on contraction behavior for a
expanding flow field with $\nabla \cdot \ \vec{v} \ \ge \ 0$ and minimal
convergence rate $|\frac{\nabla \cdot \vec{v}}{2}|$.

Note that in the case that the velocity is itself the output of a
contracting system, as e.g. will be the case later in
\ex{NavierStokes}, then the overall system constitutes a hierarchy of
contracting systems, and as such is contracting \cite{Lohm1}.

Observer designs for conservation dynamics will be illustrated in
\ex{vision} and \ex{chem}.}{conservation}

\Example{}{Consider the $m$-dimensional Bernoulli dynamics
$$ 
\frac{\partial \phi}{\partial t} + \frac{1}{2}\ \nabla \phi^T 
H^{-1}(\vec{x},t) \nabla \phi + U(\vec{x}, t) + p = 0
$$ 
with kinetic energy $\frac{1}{2}\ \nabla \phi^T H^{-1}(\vec{x},t)\nabla
\phi$, potential energy $U(\vec{x}, t)$, and $\phi$ given at all
inflowing boundary conditions of the continuum. The constraint term
$p$ (energy due to pressure) stems from the incompressibility
condition 
$$
\nabla_H \cdot H^{-1} \nabla \phi = 0
$$ 
where $\nabla_H$ denotes the covariant derivative with respect to the
metric $H(\vec{x},t)$. This dynamics governs many physical phenomena
in fluids, structural mechanics and electromagnetism. Since
$$
-\frac{\partial h}{\partial \phi}\ +\ \frac{1}{2}\ \nabla_H \cdot
\frac{\partial h}{\partial \nabla \phi}\ = 0
$$
the Bernoulli dynamics is indifferent.}{Bernoulli}

\Example{}{Approximation of continuous distributed processes is
routinely performed for numerical simulation, as well as in machine
learning and vision. Let us now briefly discuss approximating the
state of a contracting system (\ref{eq:firstorderdynamics}) as
$\hat{\Phi}(\vec{x}, t) = {\bf w}(\vec{x}, t) {\bf a}$, using $n$ basis
functions ${\bf w}(\vec{x}, t)$ and their coefficient vector ${\bf
a}$. The approximation leads to
\begin{equation} 
\frac{\partial \hat{\Phi}}{\partial t} + {\bf h} (\hat{\Phi}, 
\nabla \hat{\Phi},\vec{x}, t ) + {\bf p} = {\bf 0}
\label{eq:discretizedreaction} 
\end{equation}
where the constraint terms ${\bf p}(\hat{\Phi}, \nabla \hat{\Phi},
\vec{x}, t)$ refers to the error introduced by the approximation.  Multiplying
(\ref{eq:discretizedreaction}) from the left with ${\bf w}^T$ and integrating
results in an udpate law for ${\bf a}$,
\begin{equation}
\int_V {\bf w}^T {\bf w} \ dV \ \dot{\bf a} \ + \ \int_V {\bf w}^T \left( {\bf
h} ( {\bf w} {\bf a}, \nabla {\bf w} \ {\bf a}, \vec{x}, t ) + \frac{\partial
{\bf w}}{\partial t}\ {\bf a} \right) \ dV = {\bf 0}
\label{eq:dota}
\end{equation}
where we used the orthogonality condition $\int_V {\bf w}^T {\bf p}
dV= {\bf 0}$. The constraint terms can be computed by solving
(\ref{eq:discretizedreaction}) analytically with $\hat{\Phi}(\vec{x},
t) = {\bf w}(\vec{x}, t) {\bf a}\ $ and (\ref{eq:dota}). 

While the above is standard, the key remark is that, since a basis
function approximation $\ \hat{\Phi}(\vec{x}, t) = {\bf w}(\vec{x}, t)
{\bf a}\ $ can be interpreted as a {\it linear constraint},
contraction is {\it preserved} by the approximation (using the basic
result in~\cite{Lohm1} on adding linear constraints to contracting
systems). Note that the approximation technique can be extended to
more general partial differential equations, with the same
argumentation on the contraction behavior.

Also note that we can smoothly extend the basis functions set with
additional basis functions $w_i(\vec{x}, t)$ at any time during the
simulation, by initializing the corresponding coefficient to $a_i =
0$. Conversely, the deletion of a basis function $w_i(\vec{x}, t)$
from the set will lead to an additional disturbance $w_i(\vec{x}, t)
a_i$ and can hence only be accepted for small $|a_i|$.}{Approximation}

Finally, note that Theorem \ref{th:firstorderpde} can be extended to
the following cases:
\begin{itemize}
\item Consider a spatial discretization of $V$. Then Theorem
\ref{th:firstorderpde} can still be applied with a corresponding
discretized $\nabla$-operator since the method of proof in the
Appendix is based on such a discretization.

\item Assume that the $\nabla$-operator is generalized from a Jacobian to a
covariant derivative with respect to a symmetric metric ${\bf M}(\vec{x},
t)$. Within a local normal coordinate system \cite{Aris} defined by the metric,
the definition of $\nabla$ in (\ref{eq:firstorderdynamics}) is unchanged
(vanishing Christoffel term), so that Theorem \ref{th:firstorderpde} can be
applied unchanged.
\end{itemize}

\subsection{First-order p.d.e.'s along flow lines}
Whereas the convergence of $\phi$ is important for p.d.e.`s on a fixed
ontinuum the convergence of $\vec{x}$ is important when we consider
the dynamics along flow lines e.g. for mechanical systems or optimal control.

To assess the contraction behavior of $\vec{x}$ let us compute the
evolution of the Hessian of $\phi$ along the flow lines. We will see
later on in the optimal observer and controller dynamics and for
physical Hamiltonian systems that the definiteness of $\nabla \nabla
\phi$ will play an important role in the contraction behavior of the
ordinary plant dynamics, using \cite{Lohm1}. 

Let us compute the first and second gradient of (\ref{eq:firstorderdynamics})
\begin{eqnarray}
\frac{\partial \nabla \phi}{\partial t} &=& -\frac{\partial h}{\partial
\vec{x}} \ - \ \frac{\partial h}{\partial \nabla \phi} \nabla \nabla \phi
 \nonumber \\ \frac{\partial \nabla \nabla \phi}{\partial t} &=&
- \frac{\partial^2 h}{\partial \vec{x}^2} \ - \ \frac{\partial^2 h}{\partial
\vec{x} \partial \nabla \phi} \nabla \nabla \phi - \nabla \nabla \phi^T
\frac{\partial^2 h}{\partial \nabla \phi \partial \vec{x}} \ - \ \nabla \nabla
\phi^T \frac{\partial^2 h}{\partial \nabla \phi^2} \nabla \nabla \phi -
\frac{\partial h}{\partial \nabla \phi} \nabla \nabla \nabla \phi \nonumber
\end{eqnarray}
Following system trajectories 
\begin{equation}
\dot{\bf x} = \frac{\partial h}{\partial \nabla \phi} \label{eq:dotx}
\end{equation} leads to
\begin{eqnarray}
 \frac{d \nabla \phi}{d t} & = & \frac{\partial \nabla \phi}{\partial t} +
  \frac{\partial \nabla \phi}{\partial \vec{x}} \frac{d \vec{x}}{d t} \ = \ -
  \frac{\partial h}{\partial \vec{x}} \label{eq:dotnablaphi} \\ \frac{d \nabla
  \nabla \phi}{d t} & = & \frac{\partial \nabla \nabla \phi}{\partial t} +
  \frac{\partial \nabla \nabla \phi}{\partial \vec{x}} \frac{d \vec{x}}{d t}
  \nonumber \\ &=& \ - \frac{\partial^2 h}{\partial \vec{x}^2} \ - \
  \frac{\partial^2 h}{\partial \vec{x} \partial \nabla \phi} \nabla \nabla
  \phi - \nabla \nabla \phi^T \frac{\partial^2 h}{\partial \nabla \phi
  \partial \vec{x}} \ - \ \nabla \nabla \phi^T \frac{\partial^2 h}{\partial
  \nabla \phi^2} \nabla \nabla \phi \label{eq:Riccati}
\end{eqnarray}
where the above is known as the ordinary momentum dynamics of a Hamiltonian
(see e.g. \cite{Evan}). Equation (\ref{eq:Riccati}) generalizes the Riccati
equation (see e.g. \cite{Bryson}) to nonlinear Hamiltonian systems. 

Let us perform a modal decomposition of $\nabla \nabla \phi = \bf{X}^T
\Lambda {\bf X}$ with real eigenvalue matrix $\Lambda$ and orthonormal
eigenvector matrix ${\bf X}$. Equation (\ref{eq:Riccati}) then takes the
form
$$
 \frac{d \Lambda}{d t} = \ - {\bf X} \frac{\partial^2 h}{\partial
 \vec{x}^2} {\bf X}^T \ - \ {\bf X} \left( \frac{\partial^2
 h}{\partial \vec{x} \partial \nabla \phi} {\bf X}^T +
 \dot{\bf X}^T \right) \Lambda - \Lambda \left( {\bf X}
 \frac{\partial^2 h}{\partial \nabla \phi \partial \vec{x}}
 + \dot{\bf X} \right) {\bf X}^T \ - \ \Lambda {\bf X} \frac{\partial^2
 h}{\partial \nabla \phi^2} {\bf X}^T \Lambda
$$
where the diagonal corresponds to $n$ decoupled scalar Riccati
equations of the eigenvalues $\lambda$ in $\Lambda$.

Uniform positive definiteness of $\lambda$ (i.e. $\nabla \nabla \phi$
or convexity of $\phi$) can be shown $\forall t \ge 0$ with given
initial uniformly positive definite $\lambda_o$ or $\nabla \nabla
\phi_o$ if $\lambda$ moves away from $0$ in the neighborhood $\lambda
= 0$ and always stays upper bounded. 

Using the modular Riccati equation the first condition (moves away
from $0$) is verified if the diagonal elements
\begin{equation} 
\frac{d^j \Lambda}{dt^j}_{\Lambda = 0} = diag \left( {\bf X} L^j
\frac{\partial^2 h}{\partial \vec{x}^2} {\bf X}^T \right) \label{eq:ldhdx}
\end{equation}
are larger or equal to $0$ for $j=1$ and become uniformly positive in
one of the higher derivatives ($j>1$) if the above diagonal elements are
not uniformly positive already.

This is equivalent to requiring that $L^1 \frac{\partial^2 h}{\partial
\vec{x}^2}$ is positve semi-definite and the remaining nullspace
becomes uniformly positive in one of the higher derivatives ($j>1$)
before it eventually becomes negative. The Lie-derivatives are here defined
as
\begin{eqnarray}
L^o \frac{\partial^2 h}{\partial \vec{x}^2} &=& - \frac{\partial^2 h}{\partial
\vec{x}^2} \nonumber \\ L^{j+1} \frac{\partial^2 h}{\partial \vec{x}^2} &=&
- \frac{d}{dt} \left( L^j \frac{\partial^2 h}{\partial \vec{x}^2} \right) -
\frac{\partial^2 h}{\partial \vec{x} \partial \nabla \phi} L^j
\frac{\partial^2 h}{\partial \vec{x}^2} - L^j \frac{\partial^2 h}{\partial
\vec{x}^2} \frac{\partial^2 h} {\partial \nabla \phi \partial \vec{x}}
\nonumber
\end{eqnarray}

The inverse Riccati dynamics is given as
\begin{equation}
 \frac{d
\nabla \nabla \phi^{-1}}{dt} = \nabla \nabla \phi^{-1}
\frac{\partial^2 h}{\partial \vec{x}^2} \nabla \nabla \phi^{-1} \ + \
\nabla \nabla \phi^{-1} \frac{\partial^2 h}{\partial \vec{x} \partial
\nabla \phi} + \frac{\partial^2 h}{\partial \nabla \phi \partial
\vec{x}} \nabla \nabla \phi^{-1} \ + \ \frac{\partial^2 h}{\partial
\nabla \phi^2} \label{eq:inverseRiccati}
\end{equation}
Performing similarly to the above a modular decomposition, the second 
condition (boundedness) is verified if the diagonal elements
\begin{equation} 
\frac{d^j \Lambda^{-1}}{dt^j}_{\Lambda = 0} = diag \left(
{\bf X}^T L^j
\frac{\partial^2 h}{\partial \nabla \phi^2} {\bf X} \right) \label{eq:ldhdphi}
\end{equation}
are larger or equal to $0$ for $j=1$ and become uniformly positive in
one of the higher derivatives ($j>1$) if the above diagonal elements are
not uniformly positive already.

This is equivalent to requiring that $L^1 \frac{\partial^2 h}{\partial
\nabla \phi^2}$ is positive semi-definite and the remaining nullspace
becomes uniformly positive in one of the higher derivatives ($j>1$)
before it eventually becomes negative. The Lie-derivatives are here
defined as
\begin{eqnarray}
L^o \frac{\partial^2 h}{\partial \nabla \phi^2} &=& \frac{\partial^2
h}{\partial \nabla \phi^2} \nonumber \\ L^{j+1} \frac{\partial^2
h}{\partial \nabla \phi^2} &=& \frac{d}{dt} \left( L^j
\frac{\partial^2 h}{\partial \nabla \phi^2} \right) + \frac{\partial
h}{\partial \nabla \phi \partial \vec{x}} L^j \frac{\partial^2
h}{\partial \nabla \phi^2} + L^j \frac{\partial^2 h}{\partial \nabla
\phi^2} \frac{\partial h}{\partial \vec{x} \partial \nabla \phi}
\nonumber
\end{eqnarray}

The following theorem summarizes
this conservation law of the curvature of $\phi$ along system trajectories.

\begin{theorem}\label{Riccati-theorem}
Consider the dynamics 
$$
\frac{\partial{\phi}}{\partial t} + h (\nabla \phi, \vec{x}, t) = 0
$$ with time $t$, $m$-dimensional coordinates $\vec{x}$, and action
$\phi$.  Along trajectories, which move with flow velocity $\ \dot{\bf x}
= \frac{\partial h}{\partial \nabla \phi} \ $, the Hessian $\nabla \nabla
\phi$ evolves as
$$
\frac{d \nabla \nabla \phi}{d t} = \ - \frac{\partial^2 h}{\partial \vec{x}^2}
\ - \ \frac{\partial^2 h}{\partial \vec{x} \partial \nabla \phi} \nabla \nabla
\phi - \nabla \nabla \phi^T \frac{\partial^2 h}{\partial \nabla \phi \partial
\vec{x}} \ - \ \nabla \nabla \phi^T \frac{\partial^2 h}{\partial \nabla \phi^2}
\nabla \nabla \phi 
$$
An initial convex $\phi$ (i.e. $\nabla \nabla \phi_o > 0$) remains convex if
conditions (\ref{eq:ldhdx}) and (\ref{eq:ldhdphi}) are fulfilled.
\end{theorem}
Note that the conditions generalize the standard controllability
Grammian condition \cite{Jazwinski} to general nonlinear Hamiltonian
systems. Furthermore, they are explicitly computable, in contrast to
the controllability Grammian which requires the unknown transition
matrix.

\Example{}{\noindent Consider the plant
$$
\frac{d \vec{x}}{d t} = \vec{f}(\vec{x}, \vec{u}, t)
$$
with $m$-dimensional state $\vec{x}$, $p$-dimensional control input
$\vec{u}$, and cost-to-go dynamics
\begin{equation}
\frac{d \phi}{d t} = -\ell(\vec{x}, \vec{u}, t) \le 0 \label{eq:phidotl}
\end{equation}
along the plant trajectory with final cost $\phi_f(\vec{x})$ at time
$t_f$. A typical example is $\ \ell = \frac{1}{2}
(\vec{x}-\vec{x}_d)^T R (\vec{x}-\vec{x}_d) + \frac{1}{2}
(\vec{u}-\vec{u}_d)^T Q (\vec{u}-\vec{u}_d)\ $, with desired
trajectory $\vec{x}_d(t)$, supporting open-loop input $\vec{u}_d$, and
constant uniformly positive definite $R$ and $Q$. $\phi$ can be
minimized locally by minimizing the Hamiltonian
$$
h = \ell(\vec{x}, \vec{u}, t) + \nabla \phi \ \vec{f}(\vec{x}, \vec{u}, t)
$$
over $\vec{u}$ leading to $\frac{\partial \ell}{\partial \vec{u}} = \nabla
\phi \ \frac{\partial \vec{f}}{\partial \vec{u}}$. This leads to the
well known Hamilton-Jacobi-Bellman equation \cite{Bryson}
\begin{eqnarray}
\frac{\partial \phi}{\partial t} + h(\nabla \phi, \vec{x}, t) &=& 0
\nonumber \\ \frac{d \vec{x}}{d t} &=&  \frac{\partial h}{\partial
\nabla \phi} \nonumber
\end{eqnarray}
on a continuum $V$ which moves with $\frac{d \vec{x}}{d t}$. This equation is
solved backwards in time from final time $t_f$. Assuming $\phi_f(t_f)$ to be
convex Theorem \ref{Riccati-theorem} allows to compute under which condition
$\phi$ stays convex $\forall t < t_f$.  As a result Theorem
\ref{Riccati-theorem} allows to conclude when a global minium of $\phi$ is
given by (\ref{eq:phidotl}).

The ordinary contraction behavior of this optimal control design can be
computed with the variation of (\ref{eq:dotx})
$$
\delta \dot{\vec{x}} =  \left( \frac{\partial^2 h}{\partial
\nabla \phi^2} \nabla \nabla \phi + \frac{\partial^2 h}{\partial
\nabla \phi \partial \vec{x}} \right) \delta \vec{x}
$$
and using (\ref{eq:Riccati}) and (\ref{eq:ldhdphi})
$$
\frac{d}{dt} \left( \delta \vec{x}^T \nabla \nabla \phi \vec{x} \right)
\ = \ - \ \delta \vec{x}^T \left( \frac{\partial^2 h}{\partial \vec{x}} -
\nabla \nabla \phi \frac{\partial^2 h}{\partial \nabla \phi^2} \nabla
\nabla \phi \right) \delta \vec{x} \le \delta \vec{x}^T
L^1 \frac{\partial^2 h}{\partial \vec{x}} \delta \vec{x}
$$
We can conclude on (semi)-contraction behavior (i.e. uniformly positive
definite $\nabla \nabla \phi$ and negative (semi)-definite $L^1
\frac{\partial^2 h}{\partial \vec{x}}$) if the conditions in (\ref{eq:ldhdx})
and (\ref{eq:ldhdphi}) are fulfilled for $j = 1$ ($j \ge 1$). Note that
asymptotic convergence can even be guaranteed for semi-definite $L^1
\frac{\partial^2 h}{\partial \vec{x}}$ since the differential length cannot
get stuck due to the higher Lie derivatives in (\ref{eq:ldhdx}).

In the case that the Hamiltonian dynamics is solved numerically,
e.g. with basis functions or a neuronal network the above
argumentation stays unchanged if the Hamiltonian is extended with the
constraint term in (\ref{eq:discretizedreaction}).

If $u$ cannot be computed analytically, but rather numerically through
a convex minimization process, then the overall dynamics is still
indifferent, since it then corresponds to a hierarchy \cite{Lohm1}
composed of a contracting convex minimization process and an
indifferent Hamiltonian dynamics.}{optimalcontrol}

\Example{}{Consider again the plant
$$
\frac{d \vec{x}}{d t} = \vec{f}(\vec{x}, \vec{u}, t)
$$
with $m$-dimensional state $\vec{x}$, $p$-dimensional measurement feedback
$\vec{u}$, and cost dynamics 
$$
\frac{d \phi}{d t} = \ell(\vec{x}, \vec{u}, t) \ge 0
$$
with initial cost $\phi_o(\vec{x})$ at time $t_o$. A
typical example is $\ \ell =  \frac{1}{2} (\vec{y}-\vec{y}_m)^T R
(\vec{y}-\vec{y}_m) + \frac{1}{2}\ \vec{u}^T Q \vec{u} \ $, with measurement
$\vec{y}_m$, measurement estimate $\vec{y}(\vec{x},t)$, and constant uniformly
positive definite $R$ and $Q$. $\phi$ can be minimized by minimizing a
slightly different Hamiltonian
$$
h = -\ell(\vec{x}, \vec{u}, t) + \nabla \phi \ \vec{f}(\vec{x}, \vec{u}, t)
$$
over $\vec{u}$, leading to $\frac{\partial \ell}{\partial \vec{u}} = \nabla
  \phi \ \frac{\partial \vec{f}}{\partial \vec{u}}$. This leads again to the
  Hamilton-Jacobi equation
\begin{equation}
\frac{\partial \phi}{\partial t} + h(\nabla \phi, \vec{x}, t) = 0
\label{eq:observer} 
\end{equation}
which has to be integrated now with $t$.
This is a generalization of the well-known Hamilton-Jacobi controller design to
nonlinear optimal observer design, with the optimal solution obtained
at the minimum, where $\nabla \phi = \vec{0}$.

The convexity of $\phi$ and hence the global uniqueness of the mimimum of
$\phi$ can be shown with Theorem \ref{Riccati-theorem}. This result also hold
for an approximated system if the constraint term is added to the Hamiltonian.

Les us now show the relation of this observer design to the well-known
Kalman observer (see e.g. \cite{Jazwinski}) for linear time-varying systems.
The optimal state estimate $\hat{x}$ can the be found at the
minimum of $\phi$, i.e. for $\nabla \phi = \vec{0}$, leading with
(\ref{eq:observer}) to
$$
\frac{d \nabla \phi}{d t} = \frac{\partial \nabla \phi}{\partial t} +
  \frac{\partial \nabla \phi}{\partial \vec{x}} \frac{d \hat{x}}{d t} \ = \ -
  \frac{\partial h}{\partial \hat{x}} \ - \ \nabla \nabla \phi \frac{\partial
  h}{\partial \nabla \phi} + \nabla \nabla \phi \frac{d \hat{x}}{d t} =
  \vec{0}
$$
This yields the following observer design in $\hat{x}$ 
$$
\frac{d \hat{x}}{d t}\ = \ \frac{\partial h}{\partial \nabla \phi} \ + \
  (\nabla \nabla \phi)^{-1} \frac{\partial h}{\partial \vec{x}} 
$$ with $\nabla \phi = 0$ and the inverse covariance matrix (or
information matrix) $\nabla \nabla \phi$ defined in Theorem
\ref{Riccati-theorem}.}{optimalobserver}

\section{Second-order partial differential equations} \label{higherorderpde}

This section discusses second-order nonlinear partial differential
equations, such as e.g. diffusion equations on a closed continuum $V$, of
the form
\begin{equation}
\frac{\partial{\Phi}}{\partial t} = \nabla \cdot {\bf G} (\nabla \Phi,
\vec{x}, t) \label{eq:higherorderdynamics}
\end{equation}
where $\Phi$ is the $n$-dimensional continuous state vector, $\vec{x}$ is the
$m$-dimensional vector of underlying coordinates, and $t$ is time.

In most physical and technical applications such as diffusion and
Lagrangian dynamics, the first $\nabla$ operator and the $\nabla$
operator in the argument correspond to a limit of a left and right
differentation or vice versa. Accordingly either $\Phi$ or the
projection of $\nabla \Phi$  needs to be given on $\partial V$, so that
both $\nabla$ operators can be defined there.

Interpreting (\ref{eq:higherorderdynamics}) under the above conditions as an
ordinary infinitesimal discretization in space leads to a natural existence
conditions for the solution of (\ref{eq:pde}).

Similar to section \ref{firstorderpde} consider now two neighboring
solutions of (\ref{eq:firstorderdynamics}) {\it at fixed time $t$},
and the virtual displacement $\delta \Phi$ between them, leading to the
virtual dynamics
$$
\frac{\partial}{\partial t}\ \delta \Phi = \nabla \cdot \frac{\partial {\bf
G}}{\partial \nabla \Phi}\ \delta \nabla \Phi
$$

The time-derivative of $\int_V \delta \Phi^T \delta \Phi \ dV$ 
can be computed as
$$
\frac{1}{2} \frac{d}{dt} \int_V \delta \Phi^T \delta \Phi \ dV = - \int_V \delta
\nabla \Phi^T \frac{\partial {\bf G}}{\partial \nabla \Phi}\
\delta \nabla \Phi \ dV
$$
where we have exploited that the transpose of a left-differentation
corresponds to a negative right differentation. Lower bounding $\frac{\partial
G_{ij}}{\partial \nabla_k \phi_l} \ge \Lambda_{ijkl} \ge 0$ with
$\Lambda_{ijkl} = 0$ for $ij \neq kl$ allows to conclude on semi-contraction
behavior independent of the boundary conditions.

For given $\Phi$ on $\partial V$ the minimal contraction rate can 
be computed with a Fourier expansion as
$$
\frac{1}{2} \frac{d}{dt} \int_V \delta \Phi^T \delta \Phi \ dV \ \le \ - \
\sum_{i=1}^n \sum_{j=1}^m \frac{\Lambda_{ijij}\ \pi^2}{l_i^2} \ \int_V \delta
\Phi^T \delta \Phi \ dV
$$
where $l_i$ is the maximum length of the continuum along the $i^{\rm th}$
axis. By path integration, this immediately implies that any finite
difference between two arbitrary solutions has an equivalent contraction
behavior. Combining this result with Theorem \ref{th:firstorderpde} using the
superposition principle \cite{Lohm1} leads to
\begin{theorem}

Consider the dynamics 
$$
\frac{\partial{\phi_i}}{\partial t} + h_i (\Phi,
\nabla \phi_i, \vec{x}, t) + p_i \ = \ \nabla \cdot {\bf G}_i (\nabla \Phi, \vec{x}, t) 
$$
with time $t$, $m$-dimensional coordinates $\vec{x}$, $n$-dimensional state
vector $\Phi$, $p$ constraint terms ${\bf p}$ orthogonal to $p$ linear
algebraic constraints in $\Phi$ and $\nabla \Phi$, and $\nabla$ in line $i$ defined as
the limit of a backward differentation against the velocity $\frac{\partial
  h_i}{\partial \nabla \phi_i}$.

Let $l_i$ be the diameter (maximum length) of the continuum along the
$i^{\rm th}$ axis and lower bound the positive semi-definite
$\frac{\partial G_{ij}}{\partial \nabla_k \phi_l} \ge \Lambda_{ijkl}
\ge 0$ for a (perhaps time-varying) Dirichlet condition (i.e.,
$\Phi(t)$ specified on $\partial V$) and define $\Lambda_{ijkl} = 0 $
for the remaining (perhaps time-varying) Neumann condition (i.e.,
$\nabla \Phi(t)$ specified along the normal of $\partial V$).

For given $\phi_i$ over all inflowing boundaries $\partial V$
(i.e. $\frac{\partial h_i}{\partial \nabla \phi_i} \ {\bf n} < 0$)  and
uniformly negative definite $-\frac{\partial {\bf h}}{\partial \Phi}\ +\
\frac{1}{2}\ \nabla \cdot \frac{\partial {\bf h}}{\partial \nabla \Phi}\ $ the
solution converges unique and exponentially to a single solution, independent
of the initial conditions. The convergence rate is $| \lambda_{V} \ - \
\ \sum_{i=1}^n \sum_{j=1}^m \frac{\Lambda_{ijij}\ \pi^2}{l_i^2}|$, where $\lambda_{V}$ is the
largest eigenvalue of the symmetric part of $-\frac{\partial {\bf h}}{\partial
\Phi}\ + \ \frac{1}{2}\ \nabla \cdot \frac{\partial {\bf h}}{\partial \nabla
\Phi}$.

In the autonomous case ( ${\bf h} = {\bf h}(\Phi, \nabla \Phi, \vec{x})$,
${\bf G} = {\bf G}(\nabla \Phi, \vec{x})$ and with constant boundary
conditions and algebraic constraints) and under the same conditions, the
system converges exponentially to a unique steady-state $\frac{\partial
\Phi}{\partial t} = {\bf 0}$.
\label{th:higherorderpde}
\end{theorem}

We will call the system contracting if at least one of the two above
conditions is fulfilled. We will call such systems semi-contracting if
the above matrices are only semi-definite. And finally we will call them
indifferent if the above matrices are skew-symmetric.

The method of proof implies that all the results on
contracting systems in \cite{Lohm1} and extensions from section
\ref{firstorderpde} can be extended to contracting
reaction-convection-diffusion processes, with boundary conditions acting as
additional inputs to the system. For instance, any autonomous contracting
reaction-convection-diffusion process, when subjected to boundary conditions
periodic in time, will tend exponentially to a periodic solution of the same
period. The convergence is robust to bounded or linearly increasing
disturbances.  Finally, the method of proof also implies that any regular
spatial discretization of the above process is contracting as well, when the
explicit differentaion is performed against the velocity.

Again, the following examples will illustrate that contraction behavior can
be intpreted physically as entropy producing, since $\delta \phi$ could have
been used to spend available energy instead.

\Example{}{Consider a wafer disk, whose temperature is
controlled with a continuous external light source \cite{Lohm2}. The dynamic
equations are
$$ 
\frac{\partial \phi}{\partial t} + h \left( \phi^4 - \phi_o^4 \right) = \nabla \cdot g^{\ast}(r) \ \nabla \phi  \ \ \ \, {\rm with} \ r = \| \nabla \phi \|.
$$ 
with $\phi > 0$ the wafer temperature, $\phi_o (t)$ the external
temperature, $h$ a radiation constant, and given $\phi$ at the
boundary of the disk.  Assume that $g^{\ast}\ge 0$ and $\frac{\partial
(g^{\ast} r)}{\partial r} \ge 0$ then the Jacobian
\begin{equation*}
\frac{\partial {\bf G}}{\partial \nabla \phi} = g^{\ast} \vec{I} \ + \ r \
\frac{\partial g^{\ast}}{\partial r} \ \frac{ \nabla \phi } { \| \nabla \phi
\| }\ \frac{ \nabla \phi } { \| \nabla \phi \| }^T \ge \frac{\partial
(g^{\ast} r)}{\partial r} \ \frac{\nabla \phi } { \|\nabla \phi \| }\
\frac{\nabla \phi } { \| \nabla \phi \| }^T
\end{equation*}
is positive semi-definite. A specific saturated diffusion term may be
of the form $\ g^{\ast}= {\rm tanh} (\alpha r)/ (\alpha r)$ (with
$\alpha$ a constant), for instance.

Computing $\ \ -\frac{\partial h}{\partial \phi}\ +\ \frac{1}{2}\ \nabla
\cdot \frac{\partial h}{\partial \nabla \phi}\ = - \ 4 \ h \ \phi^3\ $,
we conclude with Theorem \ref{th:higherorderpde} on contraction
behavior with minimal convergence rate $\ 4 \ h \phi^3$.

This result may be used to design observers for the system. Indeed, it
means that an open-loop (identity) observer for the system will
converges to the actual temperature distribution with minimal
convergence rate $\ 4 \ h \hat{\phi}^3 $. Similarly, assume now that
the actual temperature is measured at the boundary of the disk. Using
this measurement {\it as a boundary condition} for the observer state
increases the convergence rate according to
Theorem~\ref{th:higherorderpde}, while at the same time preserving
consistency with the actual plant (i.e., keeping the actual
temperature as a particular solution of the observer equations with
their boundary condition). Such designs are illustrated in
\cite{Lohm2} for a linear $g$.}{waferdisk}

\Example{}{Consider the viscosity solution \cite{Evan} of the
Hamilton-Jacobi equation 
$$ 
\frac{\partial \phi}{\partial t} + h(\nabla \phi, \vec{x}, t) = g \
\nabla^2 \phi
$$ 
along a continuum $V$, which moves with the velocity $\frac{\partial h}
{\partial \nabla \phi}$ and given $\phi$ or $\nabla \phi$ along the normal of
$\partial V$.  Computing
$$
-\frac{\partial h}{\partial \phi}\ +\
\frac{1}{2}\ \nabla \cdot \frac{\partial h}{\partial \nabla \phi}\ = 0
$$ 
we can conclude with Theorem \ref{th:higherorderpde} on semi-contraction for
$g \le 0$. The minimal contraction rate of the dynamics is governed solely by
the viscosity term, the geometry and the boundary conditions of the continuum
$V$.}{dissipativeHamitonJacobi}

\Example{}{Consider the $m$-dimensional Navier-Stokes equation
$$ 
\frac{\partial {\bf v}}{\partial t} +\nabla {\bf v} \ {\bf v} + \nabla
U(\vec{x}, t) + \nabla p = g \ \nabla \cdot \nabla {\bf v} 
$$ 
with velocity ${\bf v}$, potential energy $U(\vec{x}, t)$, viscosity $g$,
incompressiblity condition $\nabla \cdot {\bf v} = 0$ leading to a pressure
gradient $\nabla p$ and ${\bf v}$ given at all boundaries. Computing
$$
-\frac{\partial {\bf h}}{\partial {\bf v}}\ +\ \frac{1}{2}\ \nabla \cdot
\frac{\partial {\bf h}}{\partial \nabla {\bf v}}\ = \frac{1}{2}\ \nabla {\bf
v}
$$ 
we can conclude with Theorem \ref{th:higherorderpde} on contraction
behavior for positive definite $\nabla {\bf v} + \sum_{i=1}^m \frac{g
\pi^2}{l_i^2} {\bf I}\ $, where $l_i$ is the diameter of the continuum
in direction $i$. Accordingly, uniqueness of Navier-Stokes solutions can be
concluded for bounded $\nabla {\bf v}$.

Comparing this result with Euler dynamics (i.e. $g=0$)  we see that the
stability is increased by the viscosity term. Comparing this result with the
velocity potential $\phi$ in the Bernoulli dynamics in \ex{Bernoulli},
potential instabilities of the Navier-Stokes can only be explained with
vortices within the dynamics. This nonlinear result on a finite convergence
region generalizes the well-known stability results on continuously
differentiable infinitesimal small disturbances in \cite{Drazin}.
}{NavierStokes}

\Example{}{Consider a continuum $V$ with $\Phi$ or $\nabla \Phi$ given
along the normal of $\partial V$, and the Lagrangian
$$ 
\mathscr{L} \ = \ T(\nabla \Phi, \vec{x}, t) \ -  \ U(\Phi, \vec{x}, t)
$$ 
Assume that $\mathscr{L}$ is convex, i.e. that $\frac{\partial^2
T}{\partial \Phi^2}$ and $\ - \ \frac{\partial^2 U}{\partial \nabla \Phi^2}$
are both uniformly positive definite. Then $\mathscr{L}$ can be
minimized using the Euler-Lagrange dynamics (see e.g. \cite{Bryson})
$$ 
\frac{\partial \Phi}{\partial t} + \frac{\partial \mathscr{L}}{\partial \Phi} 
\ = \ \nabla \cdot \frac{\partial \mathscr{L}}{\partial
\nabla \Phi} 
$$ 
Computing 
$$
-\frac{\partial h}{\partial \Phi}\ +\ \frac{1}{2}\ \nabla \cdot \frac{\partial
h}{\partial \nabla \Phi} \ = \ \frac{\partial^2 U}{\partial \Phi^2}
$$ 
contraction can be concluded using Theorem \ref{th:higherorderpde}.

This result is especially useful in function approximation and image
processing, where $\ - \ U$ describes e.g. the quadratic deviation
between measured and estimated $\Phi$, and $T$ describes a nonlinear
smoothing cost, e.g. penalizing small and medium gradients but letting
through large gradients such as edges.}{Euler-Lagrange}

\Example{}{In machine vision, consider an image (grey level) $\phi$, with
$\phi$ given at all inflowing boundaries of the imaging plane, and
$\nabla \phi$ given along the normal of the remaining
boundaries. Consider the following combination of optical flow
(conservation of brightness) from \ex{conservation} and Lagrangian
dynamics from \ex{Euler-Lagrange} (see \cite{Lohm5,Lohm6}).
$$ 
\frac{\partial \phi}{\partial t} + \vec{v} \ \nabla \phi + \frac{\partial
  \mathscr{L}}{\partial \phi} = \nabla \cdot \frac{\partial \mathscr{L}}{\partial \nabla
  \phi}
$$ 
with given camera motion flow $\vec{v}(\vec{x}, t)$.
Computing 
$$
- \frac{\partial h}{\partial \phi}\ + \ \frac{1}{2}\ \nabla \cdot
\frac{\partial h}{\partial \nabla \phi}\ \ = \ - \frac{\partial^2 \mathscr{L}}{\partial
\phi^2} + \frac{\nabla \cdot \vec{v}}{2}
$$ 
we can conclude with Theorem \ref{th:higherorderpde} on contraction behavior
for positive definite $\frac{\partial^2 \mathscr{L}}{\partial \nabla \phi^2}$ and
strictly negative $-\frac{\partial^2 \mathscr{L}}{\partial \phi^2} + \frac{\nabla
\cdot \vec{v}}{2}$.}{vision}

The approach can be vastly extended by allowing for a prior {\it
differential} coordinate transformation in $\Phi$, as in \cite{Lohm1}.
Specifically, the line vector ${\bf \delta \Phi}$ between two
neighboring trajectories can also be expressed using the differential
coordinate transformation ${\bf \delta \Psi} = {\bf \Theta} \ {\bf
\delta \Phi}$ where ${\bf \Theta(\Phi}, \nabla \Phi, \vec{x}, t)$ is a square
matrix. This leads to a generalization of our earlier definition of
squared length
\begin{equation}
  \int_V \delta {\bf \Psi}^T {\bf \delta \Psi} \ dV = \int_V {\bf \delta
  \Phi}^T {\bf M}\ {\bf \delta \Phi} \ dV \label{eq:exlength}
\end{equation}
with metric ${\bf M} = {\bf \Theta}^T {\bf \Theta}$.  Note that in
general we cannot expect to find explicit new coordinates ${\bf \Psi(\Phi},
t)$, but ${\bf \delta \Psi}$ and ${\bf \delta \Psi}^T {\bf \delta \Psi}$ can
always be defined. We require ${\bf M}$ to be uniformly positive definite, so
that exponential convergence of $\delta {\bf \Psi}$ to ${\bf 0}$ also implies
exponential convergence of $\delta {\bf \Phi}$ to ${\bf 0}$. The following
example illustrates the application of Theorem
\ref{th:higherorderpde} to new coordinates $\delta \Psi$.

\Example{}{Theorem \ref{th:higherorderpde} can be applied to a
distributed version of an example studied in detail in
\cite{Lohm2}. Specifically, consider the temperature-dependent
reaction $A \rightarrow B$ in an open reaction volume $V$ of size $100
\times 100$ units
\begin{equation*}
\frac{\partial}{\partial t} \left(
\begin{array}{c}
c_A \\ T
\end{array}
\right) +
\vec{v}^T 
\left(
\begin{array}{c}
\nabla c_A \\ \nabla T
\end{array}
\right)
= 
g \nabla \cdot \left(
\begin{array}{c}
\nabla c_A \\ \nabla T
\end{array}
\right) +
\left(
\begin{array}{c}
-1 \\ -100
\end{array}
\right) e^{-\frac{E}{T}} c_A 
\end{equation*}
with $0 \le c_A \le 1$ the partial concentration of $A$, $T > 0$ the
measured temperature, $g > 0$ the diffusion constant, $E > 0$ the
specific activation energy, and $\vec{v}(\vec{x}, t)$ the velocity field within
the reaction chamber.

The reaction volume is open at the left and the right and closed at the top
and the bottom. Accordingly we assume that $\nabla \ c_A = 0\ $ and $\ \nabla
\ T = 0\ $ everywhere along the normal of the boundary, except for $\ x = 0$
where we have $c_A = 0$ and $T = 500$. For $40 \le y \le 60\ $ the chemical
$A$ is injected with concentration $c_{Af}(t)$ and temperature $T_f(t)$.
Extending\cite{Lohm2}, the observer can be defined similarly as
\begin{equation*}
\frac{\partial}{\partial t}
\left( 
\begin{array}{c}
\hat{c}_A \\
\hat{T} 
\end{array}
\right) +
\vec{v}^T 
\left(
\begin{array}{c}
\nabla \hat{c}_A \\ \nabla \hat{T}
\end{array}
\right)
= 
g \nabla \cdot
\left(
\begin{array}{c}
\nabla \hat{c}_A \\ \nabla \hat{T}
\end{array}
\right) +
\left(
\begin{array}{c}
-1 \\ -100
\end{array}
\right) e^{-\frac{E}{\hat{T}}} \hat{c}_A +
\left( 
\begin{array}{c}
k_1 \\
k_2
\end{array}
\right)
\int_T^{\hat{T}} e^{-\frac{E}{T}} dT 
\end{equation*}

An appropriate linear coordinate transformation with constant
uniformly positive metric is computed in \cite{Lohm2} for the ordinary
reaction terms, based on a linear matrix inequality. Since the
convection and diffusion terms do not change under such a linear
coordinate transformation, Theorem \ref{th:higherorderpde} implies
contraction behavior for both observer designs and the plant for $g
\ge 0$, $\nabla \cdot \vec{v} = 0$, and uniformly negative definite
projection of the Jacobian $\frac{\partial {\bf h}}{\partial \Phi}$ on
the new coordinates.

The corresponding system responses can be accessed in \cite{plot}.}{chem}

\section{Concluding Remarks} \label{concluding}

This paper focused on the contraction analysis of important physical
and engineered first and second-order distributed systems, such as
Navier-Stokes, Euler, Lagrangian and Hamilton-Jacobi dynamics.  In
principle the technique can be extended to higher-order systems by
repeating the partial integration.

\section*{Appendix: Derivation of inequality (\ref{eq:firstordershrink})}

If either $\delta \phi_i$ or $\frac{\partial h_i}{\partial
\nabla_j \phi_i}$ is not continuously differentiable, then standard
partial integration cannot be used in the derivation of inequality
(\ref{eq:firstordershrink}). The inequality can still be obtained
using a more technical proof.  Specifically, we compute the convection
term $\frac{\partial h_i}{\partial \nabla_j \phi_i}\ \nabla_j$ in
$$
\frac{\partial}{\partial t} \delta \phi_i + \frac{\partial h_i}{\partial
\Phi}\ \delta \Phi + \sum_{j} \frac{\partial h_i}{\partial \nabla_j \phi_i}\
\nabla_j \ \delta \phi_i \ = \ 0
$$
based on the limit of a spatial discretization, where $\nabla$ is
  defined against the velocity component $\frac{\partial h_i}{\partial
  \nabla_j \phi_i}\ $, and lower bound this discretization matrix with
  the diagonal matrix $\frac{\nabla_j}{2} \cdot \frac{\partial
  h_i}{\partial \nabla_j \phi_i}\ $, so that we get inequality
  (\ref{eq:firstordershrink}) again. This lower bound is obtained
  using the following computation
\begin{eqnarray}
(\frac{\partial h_i}{\partial \nabla_j \phi_i} \nabla_j - \frac{\nabla_j}{2}
  \cdot \frac{\partial h_i}{\partial \nabla_j \phi_i}) dx_j &=&\sum_k \left(
  \begin{array}{cccccc} -|\frac{\partial h_i}{\partial \nabla_j \phi_i}_k|_+ &
  |\frac{\partial h_i}{\partial \nabla_j \phi_i}_k| & -|\frac{\partial
  h_i}{\partial \nabla_j \phi_i}_k|_-
\end{array} \right)_{k, k-1..k+1} \nonumber \\ &-& \sum_k \frac{1}{2} \left( \begin{array}{ccc}  -|\frac{\partial h_i}{\partial \nabla_j
\phi_i}_k|_+ & 0 & 0 \\ 0 & |\frac{\partial h_i}{\partial \nabla_j \phi_i}_k|
& 0 \\ 0 & 0 & - |\frac{\partial h_i}{\partial \nabla_j \phi_i}_k|_-
\end{array} \right)_{k-1..k+1, k-1...k+1} \nonumber \\ &=&
\sum_k \left( \begin{array}{ccccc} \frac{1}{2}|\frac{\partial h_i}{\partial
\nabla_j \phi_i}_k|_+ & 0 & 0 \\ -|\frac{\partial h_i}{\partial \nabla_j
\phi_i}_k|_+ & \frac{1}{2}|\frac{\partial h_i}{\partial \nabla_j \phi_i}_k| &
- |\frac{\partial h_i}{\partial \nabla_j \phi_i}_k|_- \\ 0 & 0 &
\frac{1}{2}|\frac{\partial h_i}{\partial \nabla_j \phi_i}_k|_- \end{array}
\right)_{k-1..k+1, k-1..k+1} \nonumber \\ &\ge& 0 \nonumber
\end{eqnarray}
with discretization index $k$, $|\frac{\partial h_i}{\partial \nabla_j
\phi_i}_k|_+ = \rm{max} (\frac{\partial h_i}{\partial \nabla_j \phi_i}_k, 0)$,
$|\frac{\partial h_i}{\partial \nabla_j \phi_i}_k|_- = \rm{max}
(-\frac{\partial h_i}{\partial \nabla_j \phi_i}_k,0)$, and where only the
non-zero elements of the discretization matrices are shown. In the case that
$\phi_i$ is given either at $k+1$ or $k-1$, then the corresponding row and
column must be deleted, without any impact on the definiteness.

\newpage

\end{document}